\documentclass[conference]{IEEEtran}
\IEEEoverridecommandlockouts
\usepackage{cite}
\usepackage{amsmath,amssymb,amsfonts}
\usepackage{algorithmic}
\usepackage{graphicx}
\usepackage{textcomp}
\usepackage{xcolor}
\usepackage{booktabs}
\usepackage{array} 
\usepackage{multirow}
\usepackage{balance}
\usepackage{url}

\def\BibTeX{{\rm B\kern-.05em{\sc i\kern-.025em b}\kern-.08em
    T\kern-.1667em\lower.7ex\hbox{E}\kern-.125emX}}
\begin{document}

\title{Audio-to-Image Bird Species Retrieval without Audio-Image Pairs via Text Distillation

}

\author{\IEEEauthorblockN{Ilyass Moummad}
\IEEEauthorblockA{\textit{Inria, LIRMM, Université de Montpellier} \\
Montpellier, France \\
ilyass.moummad@inria.fr}
\and
\IEEEauthorblockN{Marius Miron}
\IEEEauthorblockA{\textit{Earth Species Project} \\
Spain \\
marius@earthspecies.org}
\and
\IEEEauthorblockN{Lukas Rauch}
\IEEEauthorblockA{\textit{University of Kassel} \\
Germany \\
lukas.rauch@uni-kassel.de}
\and
\IEEEauthorblockN{David Robinson}
\IEEEauthorblockA{\textit{Earth Species Project} \\
Australia \\
david@earthspecies.org}
\and
\IEEEauthorblockN{Alexis Joly}
\IEEEauthorblockA{\textit{Inria, LIRMM, Université de Montpellier} \\
Montpellier, France \\
alexis.joly@inria.fr} 
\and
\IEEEauthorblockN{Olivier Pietquin}
\IEEEauthorblockA{\textit{Earth Species Project} \\
Belgium \\
olivier@earthspecies.org}
\and
\IEEEauthorblockN{Emmanuel Chemla}
\IEEEauthorblockA{\textit{Earth Species Project} \\
France \\
emmanuel@earthspecies.org}
\and
\IEEEauthorblockN{Matthieu Geist}
\IEEEauthorblockA{\textit{Earth Species Project} \\
France \\
matthieu@earthspecies.org}
}


\maketitle

\begin{abstract}
Audio-to-image retrieval offers an interpretable alternative to audio-only classification for bioacoustic species recognition, but learning aligned audio–image representations is challenging due to the scarcity of paired audio–image data. We propose a simple and data-efficient approach that enables audio-to-image retrieval without any audio–image supervision. Our proposed method uses text as a semantic intermediary: we distill the text embedding space of a pretrained image–text model (BioCLIP-2), which encodes rich visual and taxonomic structure, into a pretrained audio–text model (BioLingual) by fine-tuning its audio encoder with a contrastive objective. This distillation transfers visually grounded semantics into the audio representation, inducing emergent alignment between audio and image embeddings without using images during training. We evaluate the resulting model on multiple bioacoustic benchmarks. The distilled audio encoder preserves audio discriminative power while substantially improving audio–text alignment on focal recordings and soundscape datasets. Most importantly, on the SSW60 benchmark, the proposed approach achieves strong audio-to-image retrieval performance exceeding baselines based on zero-shot model combinations or learned mappings between text embeddings, despite not training on paired audio–image data. These results demonstrate that indirect semantic transfer through text is sufficient to induce meaningful audio–image alignment, providing a practical solution for visually grounded species recognition in data-scarce bioacoustic settings.\footnote{Code is available at: \url{https://github.com/ilyassmoummad/audiobioclip}}
\end{abstract}


\section{Introduction}

Automated species recognition from sensory data is a core problem in biodiversity monitoring, ecological research, and wildlife conservation. In recent years, visual species recognition has seen rapid progress driven by large-scale annotated image datasets~\cite{lasbird, tol10m, tol200m, inat21, inat24} and powerful pretrained vision--language models~\cite{bioclip, bioclip2}. In contrast, bioacoustic species recognition remains substantially more challenging. Animal vocalizations are highly variable, often corrupted by background noise and recording conditions, and typically lack the rich spatial structure available in images~\cite{hsn, nes, per, sne, ssw, uhh, ssw60}. As a result, audio-based models are generally trained on smaller datasets and encode weaker semantic structure than their visual counterparts~\cite{biolingual}.

In this work, we focus on \emph{audio-to-image retrieval} for species recognition, where an audio recording---such as a bird vocalization---is used to retrieve representative images of the corresponding species from a visual corpus. Compared to conventional audio classification, which outputs a discrete label, audio-to-image retrieval produces visually grounded and interpretable results that are particularly useful in real-world biodiversity monitoring. Retrieved images allow human users to visually verify predictions, compare similar species, and reason about uncertain or ambiguous acoustic observations. Beyond its practical relevance, this task poses a fundamental representation learning question: can species-specific acoustic cues be embedded into a visually grounded semantic space, enabling meaningful cross-modal reasoning between sound and vision?

Audio-to-image retrieval is especially appealing from a knowledge transfer perspective. Image-based species recognition models are typically trained on orders of magnitude more data than their audio-based counterparts. For instance, BioCLIP-2 is trained on hundreds of millions of biological images paired with text, spanning hundreds of thousands of taxa across diverse biological groups~\cite{bioclip2}, whereas bioacoustic models such as BioLingual rely on millions of audio recordings covering only thousands of species, primarily limited to vocalizing taxa. As a result, vision–language models encode rich, fine-grained biological semantics that extend well beyond the species observed in audio datasets. Aligning audio representations with this visually grounded embedding space therefore offers a principled way to transfer semantic structure from a strong, data-abundant modality to a weaker, data-scarce one. However, achieving such cross-modal alignment is challenging in practice due to the scarcity of paired audio–image observations.

A key obstacle is the scarcity of paired audio--image observations. In natural settings, animal sounds are frequently recorded without visual confirmation, while images rarely include synchronized audio~\cite{ssw60}. This lack of paired data makes direct alignment between audio and image encoders difficult. Consequently, existing approaches to multimodal alignment often rely on large-scale multimodal datasets and complex training pipelines that jointly optimize multiple cross-modal objectives~\cite{imagebind, languagebind, taxabind}. While effective, these methods are computationally expensive and poorly suited to bioacoustic domains, where data availability and resources are often limited.

An alternative line of work aligns pretrained audio–text and image–text models through their textual representations~\cite{mcr}. Although promising, these approaches typically rely on explicit cross-modal losses, require access to both image and audio data during training, and involve careful balancing of multiple objectives while training, which increases training complexity.

In contrast, we propose a simple and data-efficient approach that enables audio-to-image retrieval \emph{without} paired audio--image data and \emph{without} using images during audio model training. Our key idea is to use text as a semantic intermediary. Modern vision--language models trained on biological data, such as BioCLIP-2~\cite{bioclip2}, encode rich visual and taxonomic structure in their text embedding spaces. We exploit this property by distilling the text embeddings of a pretrained image--text model into a pretrained audio--text model (BioLingual~\cite{biolingual}). Concretely, we fine-tune the audio encoder to match the BioCLIP-2 text embedding space using a contrastive distillation objective, while keeping the image--text model frozen.
\looseness=-1

Through this distillation process, the audio encoder indirectly acquires visually grounded semantic structure, resulting in \emph{emergent alignment} between audio and image representations. After training, audio recordings can be projected directly into the shared image--text embedding space, enabling audio-to-image retrieval via simple similarity search. Importantly, this alignment emerges without explicit audio--image supervision or multi-objective cross-modal training, demonstrating that strong audio--image retrieval capabilities can arise from indirect semantic transfer through text alone.

Our contributions are summarized as follows:
\begin{itemize}
\item We formulate audio-to-image retrieval as a practical and interpretable task for bioacoustic species recognition, highlighting its relevance for biodiversity monitoring.
\item We introduce a simple contrastive distillation framework that transfers visually grounded semantic structure into audio representations via text, without requiring paired audio--image data or image inputs during training.
\item We demonstrate that this approach induces emergent audio--image alignment and enables strong species-level audio-to-image retrieval, outperforming baselines based on zero-shot model combinations or text embedding mappings.
\item We show that the resulting audio model preserves audio discriminative capabilities while improving text-to-audio retrieval on both focal and soundscape recordings, on average.
\end{itemize}

\section{Related Work}

\subsection{Multimodal Models for Species Recognition}

Multimodal representation learning has become increasingly important for species recognition and biodiversity analysis~\cite{bioclip, bioclip2, biolingual, taxabind, geoplant}. Vision--language models such as BioCLIP and BioCLIP-2 are trained on large-scale biological image--text pairs and encode rich, fine-grained visual and taxonomic semantics that support downstream tasks, including species classification and retrieval. In parallel, audio--language models such as BioLingual learn semantic representations of bioacoustic signals through contrastive alignment with textual descriptions. While these models align each modality to text, their embedding spaces are not directly aligned across modalities, limiting their ability to support cross-modal retrieval tasks such as audio-to-image retrieval.

\subsection{Cross-Modal Alignment and Multimodal Embeddings}

Several works aim to align multiple modalities into a shared embedding space. ImageBind~\cite{imagebind}, LanguageBind~\cite{languagebind}, and TaxaBind~\cite{taxabind} unify modalities such as image, audio, and text by jointly training on large multimodal datasets where different modality pairs share a common anchor. These approaches achieve strong cross-modal alignment but rely on extensive paired data across modalities and complex joint training procedures. In bioacoustic settings, where paired audio--image observations are rare and data collection is costly, such requirements are often impractical.

\subsection{Audio--Image Alignment via Text}

More closely related to our work, recent approaches attempt to bridge audio and image modalities through text. Multi-modal Contrastive Retrieval (MCR)~\cite{mcr} explicitly aligns pretrained image--text and audio--text models by optimizing multiple intra- and inter-modal contrastive objectives. While effective, this framework requires access to both audio and image data during training, as well as careful balancing of multiple losses over large-scale datasets, which increases computational complexity and limits applicability in resource-constrained domains.

\subsection{Positioning of Our Approach}

In contrast to existing methods, our approach enables audio-to-image retrieval without requiring paired audio–image data, explicit cross-modal losses, or image inputs during audio model training. A key enabling factor is that the pretrained audio–text and image–text models are grounded in a shared textual vocabulary, notably species names and taxonomic descriptions, which provides a common semantic anchor across modalities.

We leverage the observation that text embedding spaces of vision–language models encode visually grounded semantic and taxonomic structure. By distilling the text embeddings of a pretrained image–text model (BioCLIP-2) into a pretrained audio–text model (BioLingual), we transfer visual semantic structure into the audio representation through a simple contrastive distillation objective. This transfer induces emergent alignment between audio and image embeddings, enabling data-efficient audio-to-image retrieval using only pretrained models and audio–text data.


\section{Method}
\label{sec:method}

\begin{figure*}[ht]
  \centering
  \includegraphics[width=1.\textwidth]{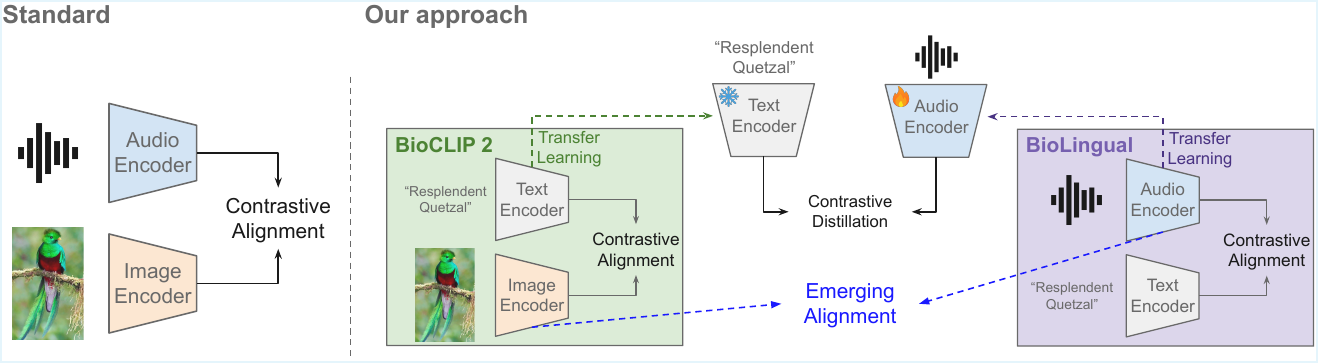}
  \caption{Comparison of audio–image alignment approaches. Left: direct contrastive alignment using paired audio–image data. Right (ours): text-based distillation from BioCLIP-2 into the BioLingual audio encoder, inducing emergent audio–image alignment without audio–image training pairs.}
  \label{fig:pipeline}
\end{figure*}

\subsection{Problem Formulation}

Let $\mathcal{A}$, $\mathcal{I}$, and $\mathcal{T}$ denote the input domains of audio recordings, images, and textual descriptions, respectively. Our goal is to learn audio representations that are aligned with image representations without audio--image pairs, enabling cross-modal tasks such as audio-to-image retrieval.

We assume access to pretrained audio–text and image–text models whose text encoders are grounded in overlapping semantic vocabularies, such as shared species names and taxonomic descriptions, though not necessarily identical label sets. This shared textual grounding provides a semantic bridge that enables indirect alignment between audio and image representations via text.

Rather than attempting to align audio and image modalities directly, we exploit the fact that modern multimodal models independently align audio and images to text. We treat text as a semantic intermediary and transfer visually grounded semantic structure from an image--text model into an audio encoder, thereby inducing indirect alignment between audio and image embeddings.
\looseness=-1

\subsection{Pretrained Multimodal Models}

\textbf{BioCLIP-2 (Image--Text).}
BioCLIP-2 is a vision--language model trained via contrastive learning on large-scale biological image--text pairs. It consists of an image encoder $f_I : \mathcal{I} \rightarrow \mathbb{R}^{d_I}$ and a text encoder $f_T^{I} : \mathcal{T} \rightarrow \mathbb{R}^{d_I}$. Textual descriptions used during training typically include taxonomic information spanning multiple biological ranks (e.g., species, genus, family), resulting in a text embedding space that captures rich visual and hierarchical biological semantics shared with the image embeddings.

\textbf{BioLingual (Audio--Text).}
BioLingual is an audio--language model trained via contrastive alignment between bioacoustic recordings and textual descriptions. It consists of an audio encoder $f_A : \mathcal{A} \rightarrow \mathbb{R}^{d_A}$ and a text encoder $f_T^{A} : \mathcal{T} \rightarrow \mathbb{R}^{d_A}$. While BioLingual encodes semantic structure relevant to bioacoustics, its embedding space is not aligned with that of BioCLIP-2, preventing direct comparison between audio and image representations.

\subsection{Visually Grounded Semantics in Text}

A key observation underlying our approach is that the text embedding space of BioCLIP-2 is not merely a label space, but a carrier of visually grounded and taxonomically structured semantics. For example, textual descriptions of the form:
\begin{quote}
\emph{``Animalia Chordata Aves Passeriformes Corvidae Pica hudsonia (black-billed magpie).''}
\end{quote}
encourage the model to organize text embeddings such that biologically related species occupy nearby regions. Because BioCLIP-2 aligns text and images in a shared space, this structure is also reflected in its image embeddings. We leverage this property to transfer visual semantic structure into the audio modality.

\subsection{Contrastive Distillation into the Audio Encoder}

To align audio representations with the BioCLIP-2 embedding space, we distill BioCLIP-2 text embeddings into the BioLingual audio encoder. Since the two models operate in different embedding dimensions ($d_A \neq d_I$), we introduce a learnable linear projection head:
\begin{equation}
g : \mathbb{R}^{d_A} \rightarrow \mathbb{R}^{d_I}.
\end{equation}

Given a batch of $N$ audio recordings $\{a_i\}_{i=1}^N$ and their associated textual descriptions $\{t_i\}_{i=1}^N$, we compute:
\begin{align}
\mathbf{z}_i^A &= g(f_A(a_i)), \\
\mathbf{z}_i^T &= f_T^{I}(t_i),
\end{align}
where $\mathbf{z}_i^A \in \mathbb{R}^{d_I}$ is the projected audio embedding and $\mathbf{z}_i^T \in \mathbb{R}^{d_I}$ is the corresponding BioCLIP-2 text embedding.

We optimize a contrastive distillation objective that encourages each audio embedding to match its corresponding BioCLIP-2 text embedding:
\begin{equation}
\mathcal{L}_{\text{distill}} =
- \frac{1}{N} \sum_{i=1}^N
\log
\frac{\exp(\mathrm{sim}(\mathbf{z}_i^A, \mathbf{z}_i^T)/\tau)}
{\sum_{j=1}^N \exp(\mathrm{sim}(\mathbf{z}_i^A, \mathbf{z}_j^T)/\tau)},
\end{equation}
where $\mathrm{sim}(\cdot,\cdot)$ denotes cosine similarity and $\tau$ is a temperature parameter.

During training, gradients are applied only to the BioLingual audio encoder $f_A$ and the projection head $g$. The text encoder of the BioCLIP-2 model is entirely frozen. This one-sided optimization transfers visually grounded semantic structure into the audio representation without modifying the image--text model or using any image data. Fig.~\ref{fig:pipeline} illustrates our distillation pipeline.
\looseness=-1

\subsection{emergent Audio--Image Alignment}

Because BioCLIP-2 aligns images and text in a shared embedding space, aligning audio embeddings to BioCLIP-2 text embeddings implicitly aligns audio and image representations. After training, semantically related audio, text, and images satisfy:
\looseness=-1
\begin{equation}
g(f_A(a)) \approx f_T^{I}(t) \approx f_I(i).
\end{equation}

As a result, audio recordings are embedded according to the visual and taxonomic semantics encoded by BioCLIP-2. Audio samples whose textual descriptions share higher-level biological attributes (e.g., order or family) are placed near images of species with the same attributes, even when fine-grained acoustic cues are weak or ambiguous. This emergent structure enables meaningful audio-to-image retrieval without explicit audio--image supervision.

\subsection{Audio-to-Image Retrieval}

\begin{figure}[h]
  \centering
  \includegraphics[width=1.\columnwidth]{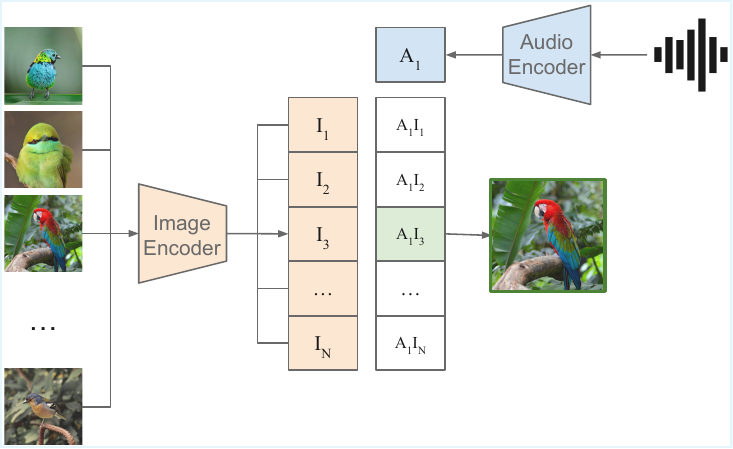}
  \caption{Audio-to-image retrieval: Images are embedded using BioCLIP-2 image encoder, while the audio query is embedded using BioLingual-FT audio encoder. Retrieval is performed by ranking images via cosine similarity in the shared embedding space.}
  \label{fig:retrieval}
\end{figure}

At inference time, text encoders are discarded. Audio recordings are embedded using the fine-tuned audio encoder followed by the projection head, while images are embedded using the frozen BioCLIP-2 image encoder. Audio-to-image retrieval is performed by ranking images according to cosine similarity with the audio query embedding, as illustrated in Fig.~\ref{fig:retrieval}.

\section{Experiments}

\subsection{Experimental Setup}

\textbf{Training protocol.}
We initialize our model from the pretrained audio encoder of BioLingual and fine-tune it using the contrastive distillation objective described in Sec.~\ref{sec:method}. During training, audio embeddings are aligned to the frozen BioCLIP-2 text embedding space. Training uses only audio–text pairs from the iNatSounds dataset~\cite{inatsounds}; no image data or paired audio–image observations are used at any stage.

For each species, textual descriptions are sampled at random from a mixture of text types supported by BioCLIP-2, including common names, scientific (Latin) names, and taxonomic descriptions (optionally combined with common names). At evaluation time, we use common-name prompts for consistency across datasets and tasks.

\textbf{Evaluation.}
We evaluate the resulting model, denoted \emph{BioLingual-FT}, along three complementary dimensions:
\begin{enumerate}
    \item emergent audio--image alignment,
    \item preservation of audio-specific representations, and
    \item quality of audio--text semantic alignment.
\end{enumerate}
Unless otherwise specified, all comparisons are against the original pretrained BioLingual model.

\begin{table}[t]
\centering
\caption{Datasets used for training and evaluation. 
ZS denotes zero-shot classification, T$\rightarrow$A text-to-audio retrieval, and A$\rightarrow$I audio-to-image retrieval.}
\label{tab:datasets}
\begin{tabular}{lccc}
\hline
\textbf{Dataset} & \textbf{\# Classes} & \textbf{Type} & \textbf{Usage} \\
\hline
iNatSounds~\cite{inatsounds} & 5,569 & Focal & Training, ZS, T$\rightarrow$A \\
SSW60~\cite{ssw60} & 60 & Focal & kNN, A$\rightarrow$I \\
HSN~\cite{hsn} & 24 & Soundscape & ZS, T$\rightarrow$A \\
NES~\cite{nes} & 51 & Soundscape & ZS, T$\rightarrow$A \\
PER~\cite{per} & 21 & Soundscape & ZS, T$\rightarrow$A \\
SNE~\cite{sne} & 20 & Soundscape & ZS, T$\rightarrow$A \\
SSW~\cite{ssw} & 35 & Soundscape & ZS, T$\rightarrow$A \\
UHH~\cite{uhh} & 24 & Soundscape & ZS, T$\rightarrow$A \\
\hline
\end{tabular}
\end{table}

\textbf{Datasets.} We evaluate our approach across multiple bioacoustic benchmarks covering both focal recordings and soundscapes. A summary of datasets and their usage is provided in Table~\ref{tab:datasets}.




\subsection{Audio-to-Image Retrieval}

We first evaluate the central task of this work: audio-to-image retrieval on SSW60. Image embeddings are computed using the frozen BioCLIP-2 image encoder, while audio queries are embedded using the fine-tuned BioLingual-FT audio encoder.

\begin{table}[h]
\centering
\caption{mean Average Precision (mAP) for audio-to-image retrieval on the SSW60 dataset.}
\label{tab:ssw60_retrieval}
\begin{tabular}{lc}
\hline
\textbf{Method} & \textbf{mAP} \\
\hline
Random Projection & 3.79 \\
Text Embeddings Mapping & 51.39 \\
Cascaded Zero-Shot (Image+Audio) & 39.85 \\
\textbf{BioLingual-FT (Ours)} & \textbf{70.47} \\
\hline
\end{tabular}
\end{table}

We compare our approach against several baselines that reflect different strategies for bridging audio and image modalities without paired audio--image supervision:

\emph{Random Projection} applies a random linear projection to BioLingual audio embeddings to match the dimensionality of BioCLIP-2 image embeddings and serves as a lower bound.

\emph{Text Embeddings Mapping} learns a non-linear mapping between the text embedding spaces of BioLingual and BioCLIP-2 via the contrastive loss using iNatSounds annotations spanning 5,569 classes. 
Text captions of iNatSounds classes are embedded with BioLingual and BioCLIP-2 text encoders and then are aligned via the contrastive loss. This baseline tests whether translating BioLingual's text embedding to BioCLIP-2's text embedding without audio is sufficient for cross-modal retrieval.
\looseness=-1

\emph{Cascaded Zero-Shot} decomposes audio-to-image retrieval into two independent zero-shot classification steps: audio is first classified into one of 3,846 bird classes from iNatSounds using BioLingual, and images are independently classified using BioCLIP-2. Retrieval is then performed by matching predicted class embeddings. This baseline reflects a common cascade approach but is susceptible to error propagation across stages.
\looseness=-1

\emph{BioLingual-FT}, trained via our proposed text-based distillation strategy, substantially outperforms baselines, despite never observing images or paired audio--image data during training. These results demonstrate that strong audio--image alignment can emerge purely through indirect semantic transfer via text, validating the effectiveness and data efficiency of the proposed method.
\looseness=-1



\subsection{Preservation of Audio Representations}

A key concern when distilling visual semantics into an audio encoder is potential degradation of audio-discriminative features. To assess this, we evaluate $k$-nearest neighbor (kNN) classification accuracy on the SSW60 dataset, a standard benchmark for fine-grained categorization of bird species.

\begin{table}[h]
\centering
\caption{kNN classification accuracy on SSW60.}
\label{tab:ssw60_knn}
\begin{tabular}{lc}
\hline
\textbf{Model} & \textbf{Accuracy} \\
\hline
BioLingual & 77.37 \\
BioLingual-FT & 77.29 \\
\hline
\end{tabular}
\end{table}

As shown in Table~\ref{tab:ssw60_knn}, BioLingual-FT matches the performance of the original BioLingual model, indicating that the proposed distillation preserves core audio representations while injecting additional semantic structure.

\subsection{Audio--Text Alignment on Focal Recordings}

We next evaluate audio--text alignment on focal recordings from iNatSounds. We report both text-to-audio retrieval performance and zero-shot classification accuracy, which directly probe the semantic consistency between audio and textual embeddings.
\looseness=-1

\begin{table}[h]
\centering
\caption{iNatSounds text-to-audio retrieval (mAP@1000) and zero-shot classification accuracy on validation and test splits.}
\label{tab:inat}
\begin{tabular}{lcccc}
\hline
\textbf{Model} 
& \multicolumn{2}{c}{\textbf{Text$\rightarrow$Audio}} 
& \multicolumn{2}{c}{\textbf{Zero-Shot}} \\
\cline{2-5}
& \textbf{Val} & \textbf{Test} & \textbf{Val} & \textbf{Test} \\
\hline
BioLingual & 60.89 & 58.77 & 38.45 & 36.29 \\
BioLingual-FT & \textbf{76.45} & \textbf{77.97} & \textbf{64.88} & \textbf{63.32} \\
\hline
\end{tabular}
\end{table}

BioLingual-FT substantially outperforms the base model, improving retrieval performance by over 19\% in mAP and nearly doubling zero-shot classification accuracy. These gains demonstrate that aligning audio embeddings to the BioCLIP-2 text space significantly strengthens semantic grounding for focal bioacoustic recordings.

A potential concern is that the strong gains observed on iNatSounds could partially stem from incidental overlap between the species vocabulary or underlying observations used to train BioCLIP-2 (on images) and the iNatSounds audio recordings. While exact audio--image overlap is highly unlikely in practice, both datasets are drawn from large biodiversity repositories and may share species-level semantic context.

To assess whether our improvements reflect genuine audio--text alignment rather than dataset-specific effects, we evaluate both the original BioLingual model and BioLingual-FT on the CBI dataset~\cite{cbi}, a focal bioacoustic dataset that is disjoint from iNatSounds and not used during training.

\begin{table}[h]
\centering
\caption{Zero-shot classification accuracy on the CBI dataset.}
\label{tab:cbi}
\begin{tabular}{lcc}
\hline
\textbf{Model} & \textbf{Zero-Shot Acc. (\%)} & \textbf{A$\rightarrow$I mAP (\%)} \\
\hline
BioLingual & 58.43 & 51.44 \\
BioLingual-FT & \textbf{58.77} & \textbf{72.30} \\
\hline
\end{tabular}
\end{table}

As shown in Table~\ref{tab:cbi}, BioLingual-FT substantially improves text-to-audio retrieval on CBI (+20.9 mAP) while preserving zero-shot classification accuracy. Because CBI is independent of iNatSounds and was not used during training, these results indicate that the audio--text improvement on focal recordings does not stem from dataset overlap.

\subsection{Audio--Text Alignment on Soundscape Recordings}

We further test generalization to real-world soundscapes, which are more challenging due to background noise and overlapping species. Table~\ref{tab:soundscapes} reports zero-shot classification accuracy and text-to-audio retrieval across six soundscape datasets.
\looseness=-1

\begin{table}[h]
\centering
\caption{Soundscape zero-shot classification accuracy and text-to-audio retrieval (mAP@1000).}
\label{tab:soundscapes}
\resizebox{\columnwidth}{!}{
\begin{tabular}{lcccc}
\hline
\multirow{2}{*}{\textbf{Dataset}} &
\multicolumn{2}{c}{\textbf{Zero-Shot (Acc.)}} &
\multicolumn{2}{c}{\textbf{Text$\rightarrow$Audio (mAP)}} \\
\cline{2-5}
 & \textbf{BioLingual} & \textbf{BioLingual-FT} & \textbf{BioLingual} & \textbf{BioLingual-FT} \\
\hline
HSN & \textbf{19.69} & 17.93 & 27.85 & \textbf{45.40} \\
NES & \textbf{32.18} & 18.46 & 24.32 & \textbf{33.50} \\
PER & \textbf{7.68} & 2.95 & \textbf{9.73} & 6.43 \\
SNE & 42.70 & \textbf{42.89} & 21.49 & \textbf{42.75} \\
SSW & 35.98 & \textbf{47.13} & 33.64 & \textbf{59.29} \\
UHH & \textbf{16.65} & 8.34 & \textbf{32.00} & 27.54 \\
\hline
AVG & 25.81 & 22.95 & 24.83 & 35.81 \\
\hline
\end{tabular}
}
\end{table}

While zero-shot classification accuracy shows mixed behavior across datasets, BioLingual-FT yields a large average mAP increase in text-to-audio-retrieval. This suggests that the distilled model prioritizes semantic alignment over strict class discrimination, a trade-off that is beneficial for retrieval-based and exploratory tasks.

\section{Conclusion}



%

We introduced a simple and data-efficient approach for audio-to-image species retrieval in bioacoustic settings where paired audio--image data are unavailable. By using text as a semantic intermediary, we distill visual--text embeddings from BioCLIP-2 into a pretrained BioLingual audio encoder, inducing direct audio--image alignment without requiring image supervision.
\looseness=-1

Experiments show that the proposed model achieves strong audio-to-image retrieval performance, substantially outperforming zero-shot and text-embedding mapping baselines. These results demonstrate that text-based distillation alone is sufficient to induce effective audio–image alignment, providing a practical and lightweight solution for visually grounded bioacoustic analysis in data-scarce biodiversity applications. Importantly, this emergent cross-modal alignment does not come at the expense of audio modeling: the distilled audio encoder preserves audio discriminative capabilities and improves text-to-audio retrieval.



\bibliographystyle{IEEEtran}
\balance
\bibliography{refs}

\end{document}